\documentclass[12pt,titlepage]{article}
\usepackage[pdftex]{graphicx}
\usepackage{subfigure}
\usepackage{mathrsfs}
\usepackage{geometry}
\usepackage[dvips]{epsfig}
\usepackage{mathrsfs}
\usepackage{amsfonts}
\usepackage{amssymb}

\long\def\symbolfootnote[#1]#2{\begingroup
       \def\thefootnote{\fnsymbol{footnote}}\footnote[#1]{#2}\endgroup}

\begin{document}
\date{}
\title{\flushright{\small Report
No.AE-A\&S-01\\November 2004}\\Morphological Analysis of Cryogenic
Spray Images}
\author{ Hrishikesh Ganu and  B.N.
Raghunandan \\Atomization and Sprays Laboratory\\ Department of
Aerospace Engineering
\\ Indian Institute of Science \\ Bangalore-560012 }
\maketitle \tableofcontents
\newpage
\section{Introduction}
This study gives the development of a new technique for analyzing
images of Cryogenic sprays, to estimate the drop-size
distribution. It has a sound mathematical basis, in the form of
Mathematical Morphology, and we have tried to build up a
formulation for a granulometry, starting from the elementary
operations of Dilation and Erosion. An axiomatic foundation for
granulometry has also been discussed.

We have taken an actual $LN_2$ spray photograph for analysis, \ to
illustrate the use of Morphological operations, culminating in a
granulometry.
\section{The Malvern Mastersizer} The Malvern
mastersizer is based on the principle of laser ensemble light
scattering.It is currently,\ the most widely used instrument for
estimating the droplet size-distributions of sprays. Ensemble
scattering means that droplets cannot be measured individually;
only a group of a large number of droplets can be sized.It is a
non-imaging instrument, since sizing is done without forming an
image of the object.  \\A laser beam emerging from the source, is
scattered by a collection of droplets in a spray, and the
scattered light falls on the receiver.The receiver is an array
having several annular detector elements.Each element senses the
light scattered by droplets in a particular size-class. The sizes
of the droplets are related to the angle of scattering.
\section{Need for
an Imaging system}In case of the Malvern Mastersizer,sufficient
light must fall on the detector for it to  sense the amount of
scattering. Excessive obscuration, encountered in dense sprays
 prevents this.\\Further, in the specific case of a liquid
 nitrogen spray, injected into the ambient at supercritical
 temperature, vapour around the evaporating droplet leads to
 density gradients, which in turn affect the refractive index.This effect is called
 beam steering.

As it is, Malvern Mastersizer gives an excessively large
SMD(typically, $800\mu m$ at 450kPa).We know from the physics of
droplet breakup, that this is not possible.Moreover, studies by
Chin et.al \cite{ch} indicate that the Malvern Mastersizer
underpredicts the SMD, of dense sprays.Thus
 the drop size distribution given by Malvern Mastersizer is
 physically untenable.
              There is, therefore,a need for an imaging system, which can
              directly capture the spray on an image, which can
              then
              supply information about the drop sizes.

\section{Method of capturing images}\
We have used $LN_2$ as a simulant for cryogenic propellants. The
axisymmetric jet issues out of the plain injector, into the
ambient at pressure drops between 300kPa and 700kPa. For capturing
the images a 1540-P2 strobolume pulse light generator together
with a digital camera of resolution 300dpi is used with
front-lighting.The width of the pulse is 50 ns at a light power
output of 250 mW. The wavelength of emission is 660 nm with a
spectral width of 27nm. Images are captured at a high film speed
equivalent to ISO 400 on a conventional photographic film.

\section{Image Processing} Photographs of the spray
yield just a raw image, which can scarcely be of any use. It must
be first pre-processed to a form on which Morphological operations
can  be performed. Pre-processing includes conversion to
grayscale, followed by contrast enhancement if required. The
grayscale version is then converted to binary, using an
appropriate threshold.These are standard image processing
operations, and details may be found in \cite{Gon}

\section{Mathematical Morphology } Morphology is a term commonly used in the
biological sciences for describing the form or structure of an
organism. In image processing , by morphology, we mean the
geometrical or textural
features of the image.\\
The techniques of morphology try to duplicate the manner in which
a human being perceives an image. While observing a flower for
example, we might first observe the colour, then the texture,
followed by the geometry and so on, in succession not all at the
same time. We then try to look at that aspect, which interests us
the most, in greater detail.\\

In morphological image processing(MIP) too, we select a specific
characteristic of interest, and analyze the image with respect to
that.Knowledge about the other characteristics is now irrelevant
to us and, is off-loaded before we proceed. This is why in our
experiments, the binary version has been used, though we could as
well have used the grayscale version.
\\
\subsection{The basic morphological operations} Here
are some basic operations which are extensively used in MIP.The
book by Gonzalez and Woods \cite{Gon} is an excellent introduction
to Morphology. Let A be a set in $\Re^2$ and $x\in \Re^2$
\begin{enumerate}
\item
\textbf{Translation} \\
$A+x=\{a+x:a\in A\}$ \\
Which is to say that a point $z\in A+x \Leftrightarrow \exists $\
a point $a\in A,$ such that z=a+x.

\item \textbf{Minkowski addition} \\
For A and B $\subset \Re^2$\\
$A\oplus B=\cup A+x,$ with $x\in B$. This leads to:\\ $A\oplus
B=\cup \{x+y \}$, such that $x\in A$ and $y\in B$. Several
properties immediately follow this definition:\\ \begin{enumerate}
\item $A+\bar{0}=A.$
\item $A\oplus x=A+x \ \forall \ x\in \Re^2$.
\item Commutativity, $A\oplus B =B\oplus A$.
\item Associativity, $A\oplus(B\oplus C)=(A\oplus B)\oplus C$, and in
particular $A\oplus[B +x]=[A\oplus B] + x $
\end{enumerate}
\item \textbf{Minkowski subtraction}\\
$A\ominus B=\cap A + x$, such that $x\in B$. With \textit{scalar
multiplication} being defined as $tA=\{tx:x\in A\}$ we get $
-A=\{-x:x\in A\}$, which is called the \textit{reflection} of A.
\item \textbf{Erosion}
An alternative expression , is:\\$A\ominus B =\{x:-B+x\subset
A\}$, from which the erosion is defined as:\\$A\ominus
-B=\mathscr{E} (A,B)= \{x:B+x\subset A\}$.
\item \textbf{Dilation}
It is the same as Minkowski addition\\
$\mathscr{D}(A,B)=A\oplus B$. \\Dilation and Erosion obey:\\
\begin{figure} [pt]
\centering
\includegraphics [scale=.65]{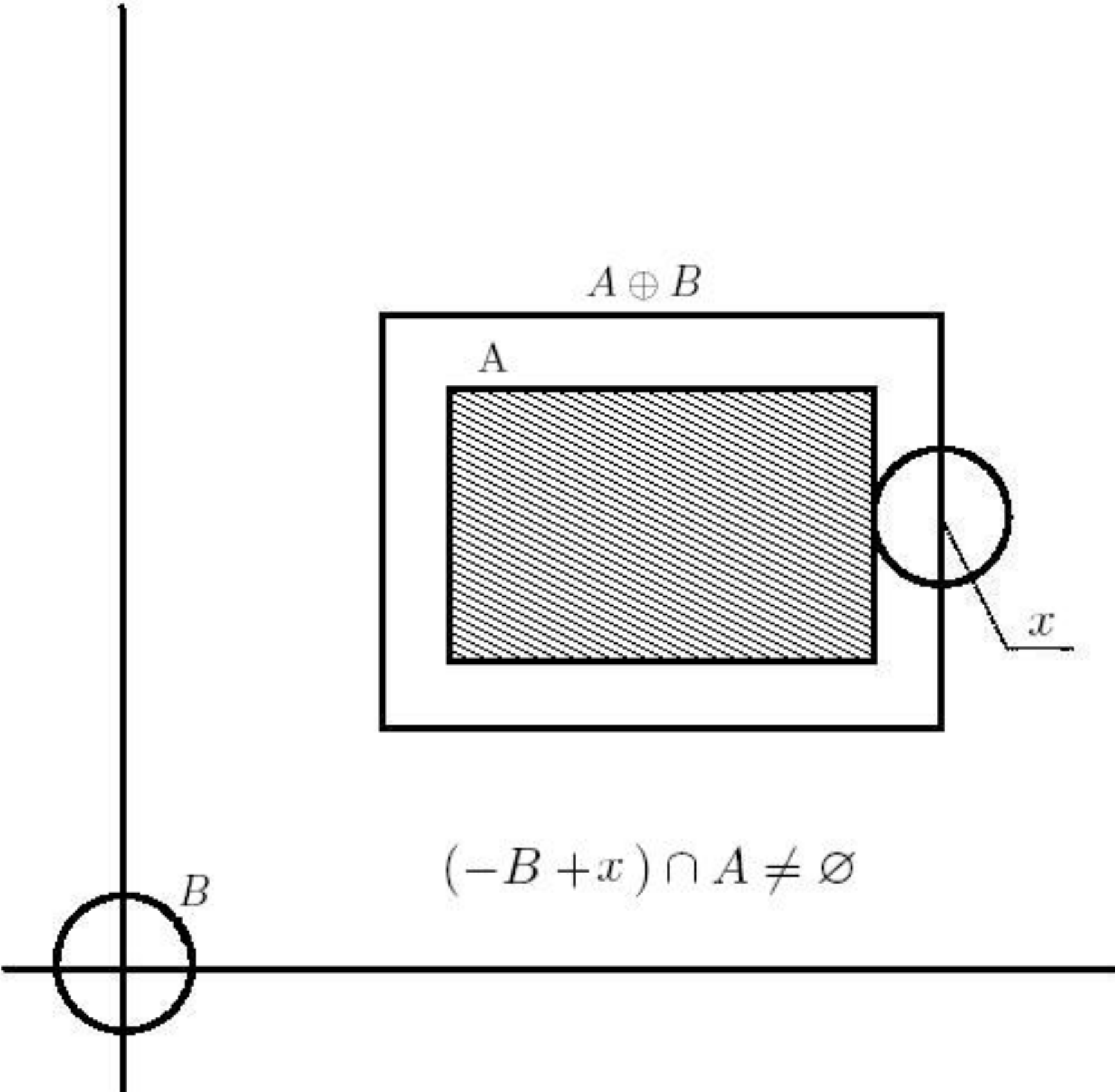}
\caption {Dilation of an image by a structuring element}
\label{dilation}
\end{figure}

\begin{enumerate}
\item $\mathscr{D}(A,B)=\mathscr{D}(B,A)$
\item $\mathscr{D}(A, B+x)=\mathscr{D}(A,B)+x.$
\item $\mathscr{E}(A,B)=\mathscr{E}{A,B+x}+x.$
\item For a given B, and $A_1\subset A_2$,
\begin{enumerate}
\item $\mathscr{D}(A_1,B)\subset \mathscr{D}(A_2,B)$
\item $\mathscr{E}(A_1,B)\subset \mathscr{E}(A_2,B).$
\end{enumerate}

\begin{figure} [hpt]
\centering
\includegraphics [scale=.65]{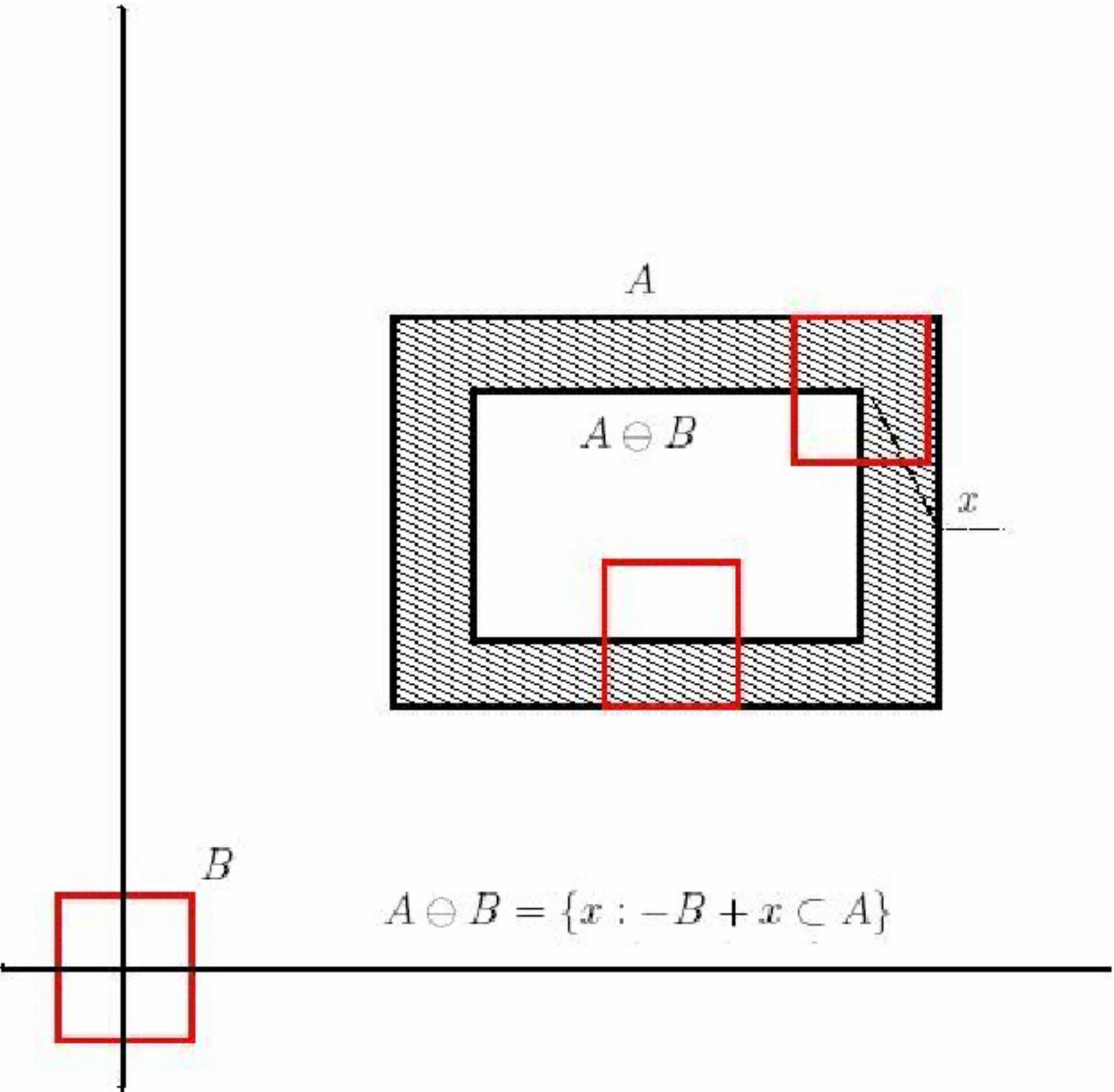}
\caption {Erosion of an image} \label{erosion}
\end{figure}

\begin{figure} [hpt]
\centering
\includegraphics [scale=.65]{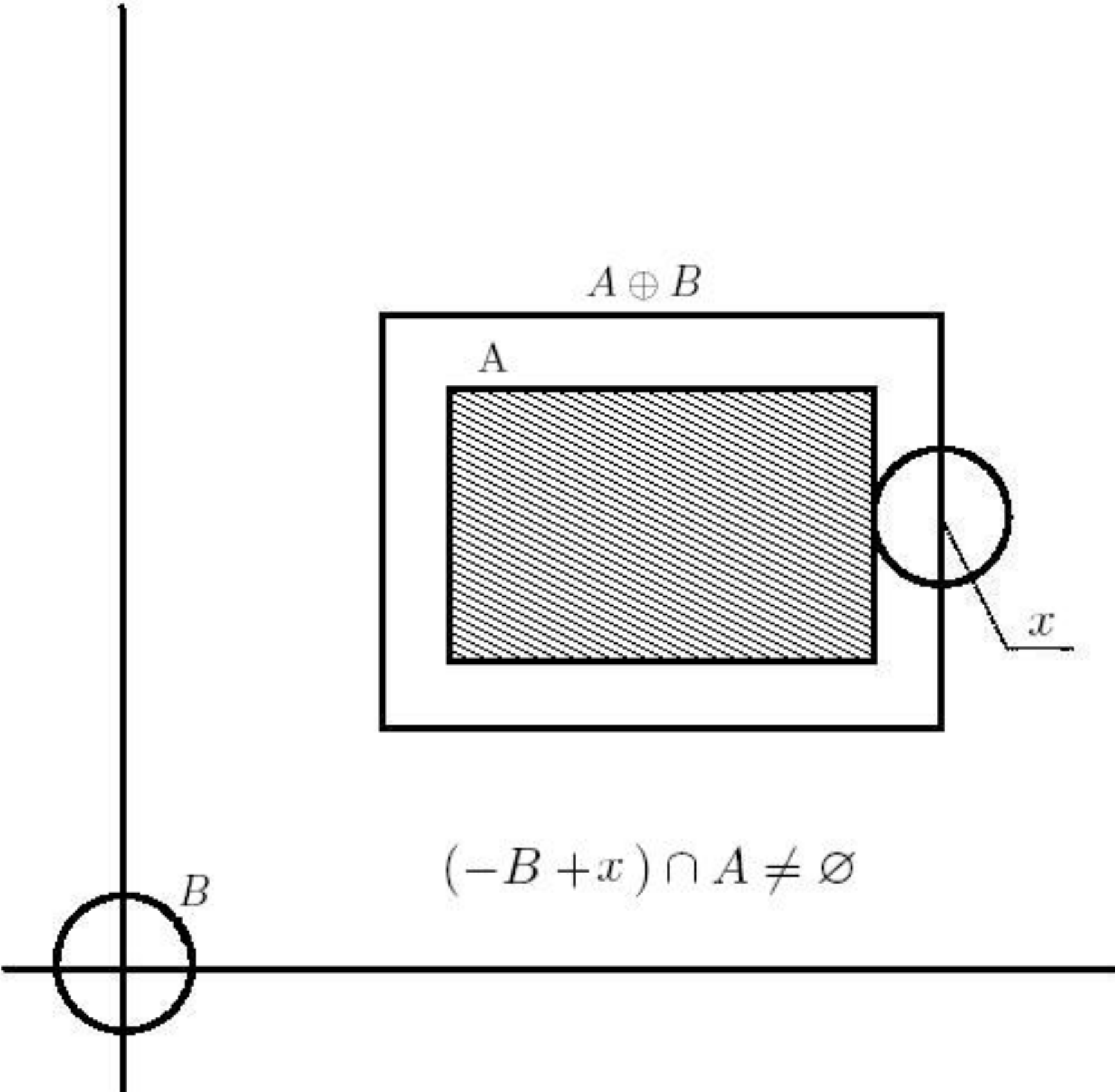}
\caption {Dilation of an image} \label{erosion}
\end{figure}

\begin{figure} [pt]
\centering
\includegraphics [scale=.65]{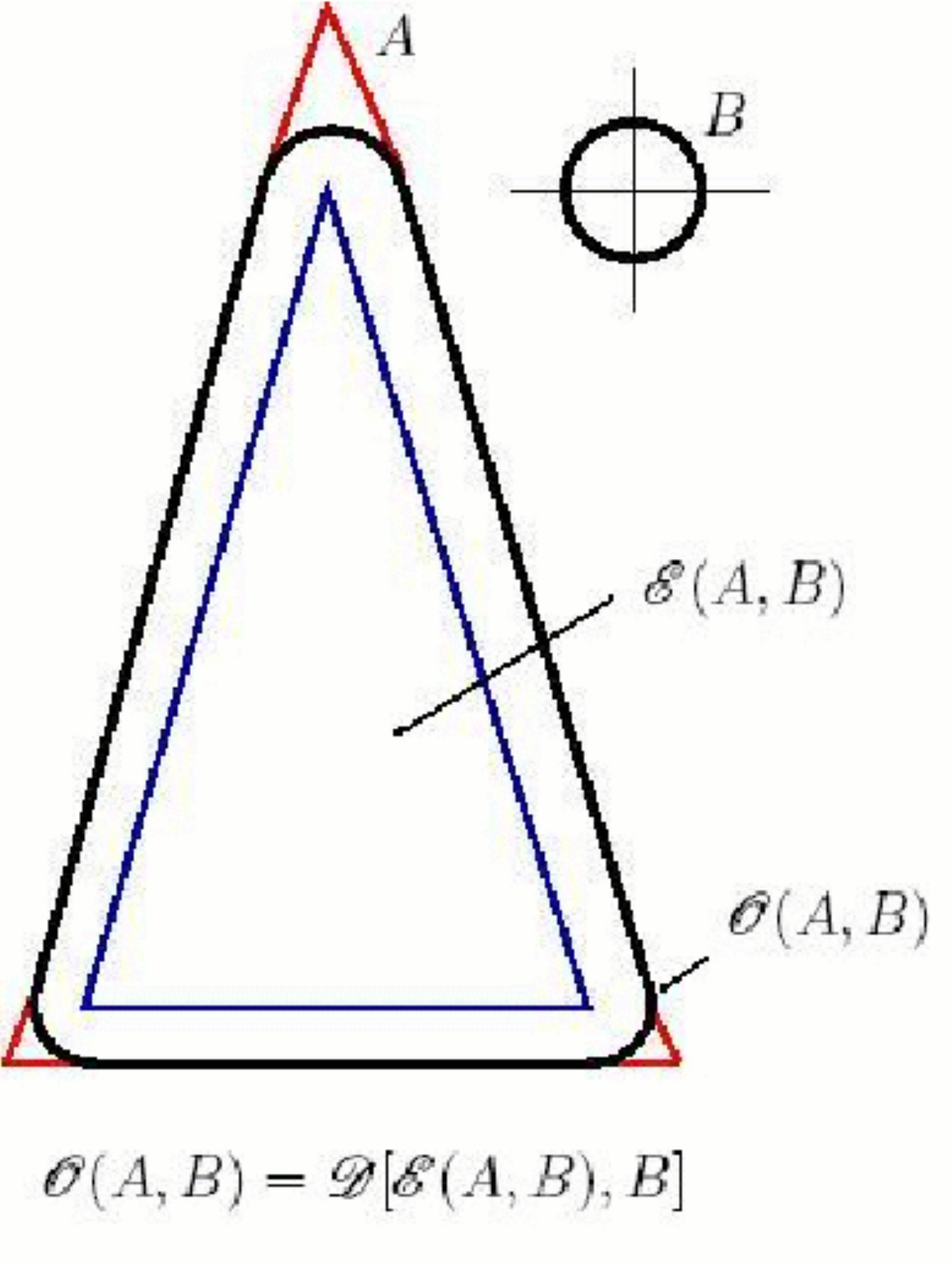}
\caption {Opening, as erosion followed by dilation}
\label{opening}
\end{figure}

\item \textbf{Opening}
Opening is a combination of dilation and erosion and is defined
as:\\
$\mathscr{O}(A,B)= \mathscr{D}[\mathscr{E}(A,B),B]$
\item \textbf{Closing}
$\mathscr{C}(A,B)= \mathscr{E}[\mathscr{D}(A,-B),-B]$.These
operations will now be analyzed in detail, since they were used to
analyze cryogenic spray photographs in this exercise.
\end{enumerate}
\end{enumerate}
\begin{figure} [pt]
\centering
\includegraphics [scale=.65]{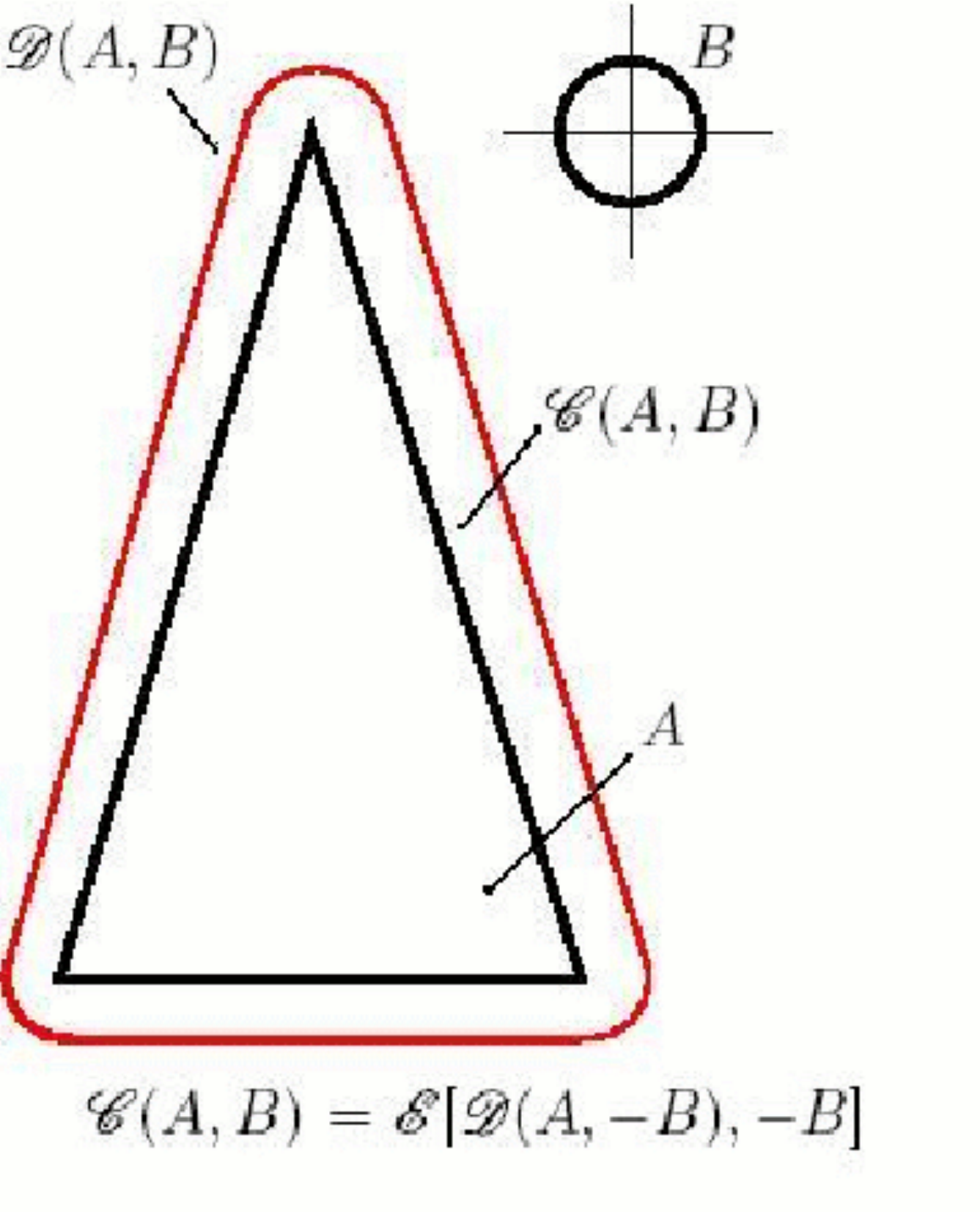}
\caption {Closing operation} \label{closing}
\end{figure}
\newpage
\subsection{Analysis of Opening and Closing}

\begin{enumerate}
\item $\mathscr{C}(A,B)^c=\mathscr{O}(A^c,B)$ and \\
\item $\mathscr{O}(A,B)^c=\mathscr{C}(A^c,B)$ \ldots Duality.\\
Thus,if the complement of the first operand, A is known, we can
find the opening, given the closing and vice-versa.
\item $\mathscr{O}(A,B)=\cup \{B+y: B+y\subset A\}$\\
which says that the opening is the union of translates of B which
are contained within A.
\item$\mathscr{C}(A,B)=\cap \{(B+y)^c: (B+y)\subset A^c\}$\\
Closing is the intersection of the complements of certain
translates of B.
\item A point z is $\in \mathscr{C}(A,B)\Leftrightarrow
(B+y) \cap A \neq \phi \ \forall (B+y)$
\item $\mathscr{O}(A,B)\subset A$\\ The opening is
antiextensive.\label{com1}
\item $A_1\subset A_2\Rightarrow \mathscr{O}(A_1,B)\subset
\mathscr{O}(A_2,B)$\\It is increasing.
\item $\mathscr{O}[\mathscr{O}(A,B),B]=\mathscr{O}(A,B)$\\
and also idempotent.\\
The closing is also increasing and idempotent,
\item $\mathscr{C}(A,B)\supset A$, \ but is extensive.
\\A set A is \textit{B-closed}(or \textit{B-open})  when
$\mathscr{C}(A,B)=A(or\mathscr{O}(A,B)=A).$
\item  A is B-closed $\Leftrightarrow A^c$ is B-open.
\item This is connected with the Minkowski, addition and subtraction as:\\
A is B-open$\Leftrightarrow  \  \exists \ $a set E such that
$A=E\oplus B$ and \\
A is B-closed$\Leftrightarrow  \  \exists \ $a set E such that
$A=E\ominus B$
\item Generalized laws of idempotency for a set A which is B-open,and any other set F:\\
$\mathscr{O}[\mathscr{O}(F,B),A]= \mathscr{O}(F,A) $\ and\\
$\mathscr{O}[\mathscr{O}(F,A),B]= \mathscr{O}(F,A)$
\item If $r\geq s\> 0$, with $r ,s,\ \in \Re$ and B is convex\symbolfootnote[1]{A
set is called convex when a line joining any two of its points,\
lies entirely within the set.}, then for any A,
$\mathscr{O}(A,rB)\subset\mathscr{O}(A,sB)$
\end{enumerate}
\newpage
\section{Granulometry}
\label{sec:this} Let A be a compact set $\subset \Re^2$
\symbolfootnote[2]{\begin{enumerate}
\item A compact set is closed and bounded.
\item A closed set contains its own boundary.
\item A bounded set is contained within a disk of finite radius.
, centred at the origin. \end{enumerate}} and t $\in \Re$. We, \
next consider $\Psi$ an image-image transformation(which could be
the opening,\ on a cryogenic spray image, for instance), \ applied
to A\\
$\Psi :A  \longmapsto \Psi(A)$ .\\
If E is a convex, \textit{generating structuring element}(which is
really, an image).Then \{tE \} is a family of structuring
elements.The mapping $t\longmapsto \mathscr{O}(A,tE)$is
decreasing.This means that$\mathscr{O}(A,tE)$ shall be empty
$^\bigstar$ for sufficiently large t.The mapping $t\longmapsto
\mathscr{O}(A,tE)$is known as a \textit{granulometry}.
Granulometric filtering is analogous to the physical operation of
sieving.Intuitively, it is sufficient to understand that this is a
kind of filtering in which those sections of the image which are
not sufficiently large to hold the structuring element tE are
removed from the image.

 Though this can be better appreciated, by
following the algorithm, with a photograph of an actual cryogenic
spray as will be seen in this sub-section,the axiomatic
development of granulometries, given in the next sub-section,
should not be skipped if one wishes to really understand the
process. This study is on Binary images;Morphological operations
in the gray-scale domain
are discussed in \cite{Har}.\\
\subsection{Sieving and  Granulometry on a Cryogenic Spray
Photograph} 
\begin{figure} [pt]
\centering
\includegraphics [scale=.65]{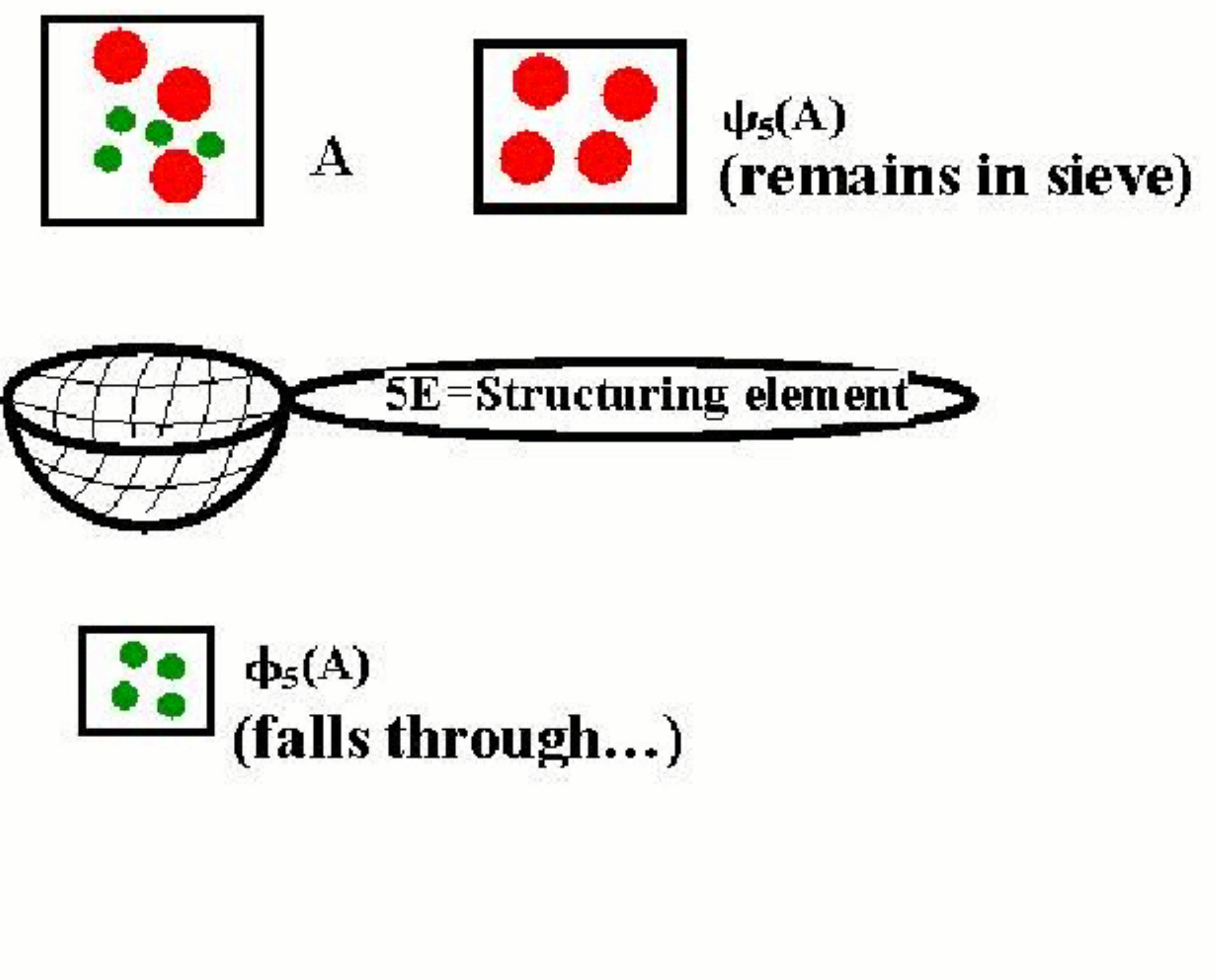}
\caption {Physical Analogy (Sieving) for Opening} \label{sieve1}
\end{figure}

The algorithm for generating a granulometry is
represented by the flowchart shown in Figure \ref{flow} below. It
accepts the raw image A as the input and delivers a single
parameter granulometry $\Phi_t(A)$\,with\ $t=1,2,\ \ldots$ as the
parameter.The \textit{generating} structuring element used,\ E, is
a unit square.\\
     Incrementing the value of t after each step yields a sequence
     of images under the granulometry $t\mapsto
     \mathscr{O}(A,tE)$. The process stops when  at some stage the opening is
     empty(as it shall be; see Section  \ref{sec:this},\ $\bigstar$)
\begin{figure} [pt]
\centering
\includegraphics [scale=.65]{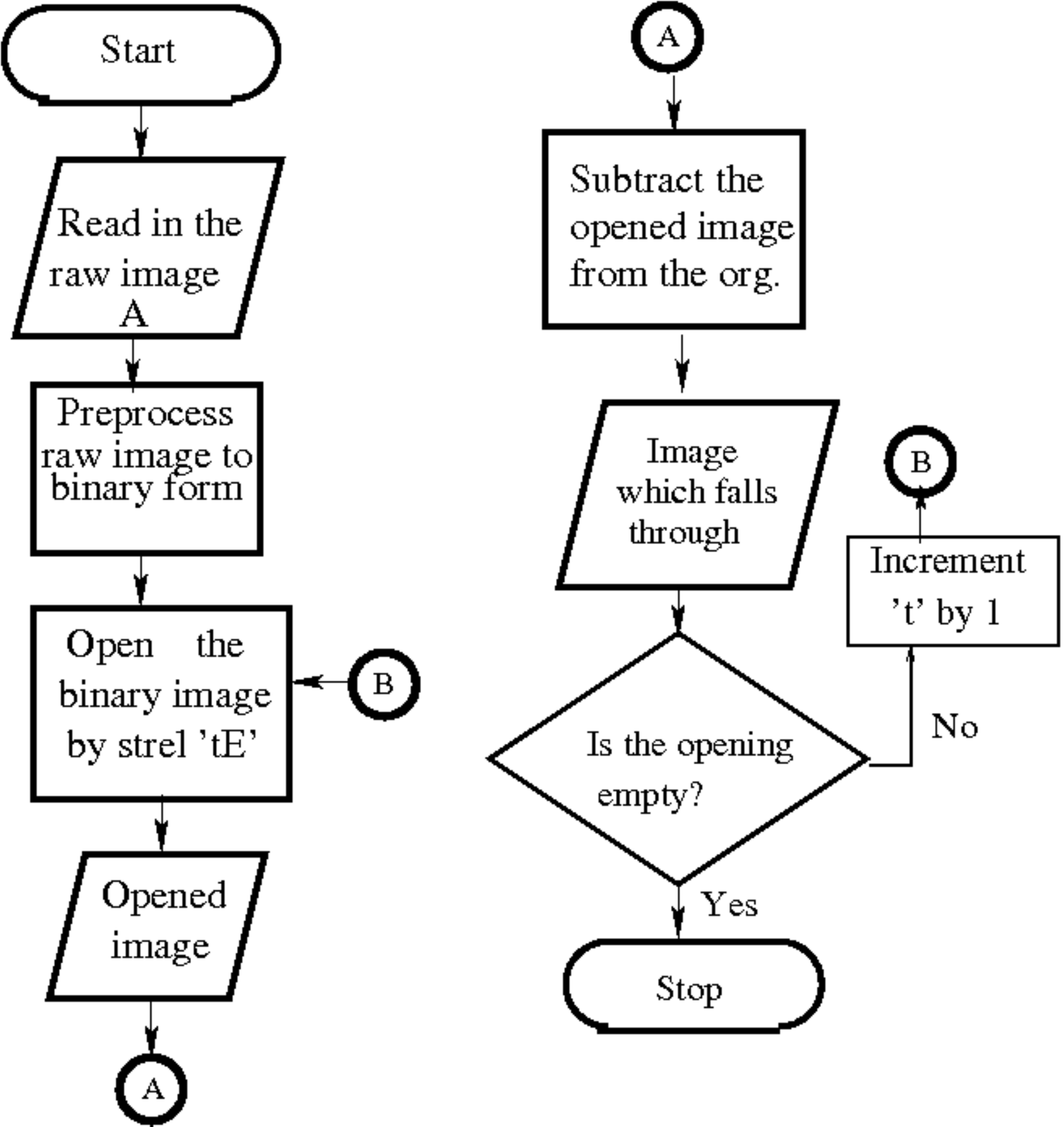}
\caption {Flowchart for generating a granulometry} \label{flow}
\end{figure}
Figure \ \ref{sieve1} shows the physical analogy for the
operation. A sequence $\Psi_t(A)$ of openings is thus produced.It
can be seen that there will be a minimum mesh size at which
nothing will be retained on the sieve.
\begin{figure} [pt]
\centering
\includegraphics [scale=.75]{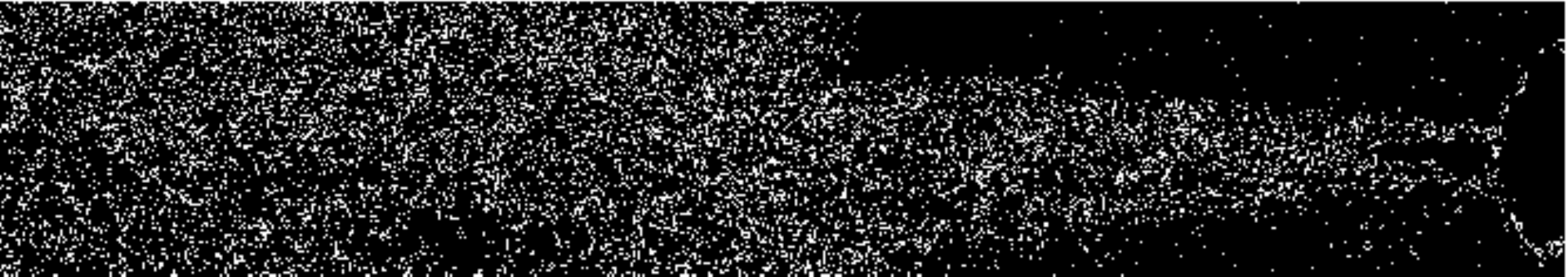}
\caption {Opening by a structuring element of size 4 units
$\Psi_4(A)$} \label{op4}
\end{figure}
\begin{figure} [pt]
\centering
\includegraphics [scale=.75]{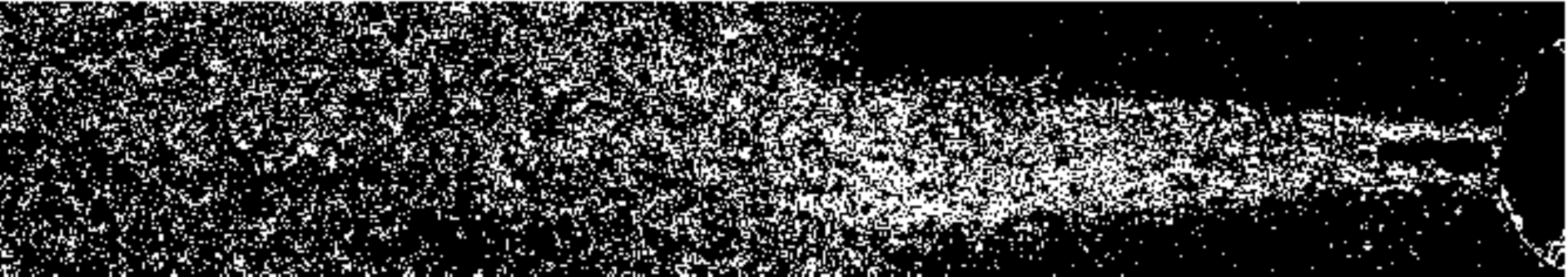}
\caption {Opening by a structuring element of size 8 units
$\Psi_8(A)$} \label{op8}
\end{figure}

\begin{figure} [pt]
\centering
\includegraphics [scale=.75]{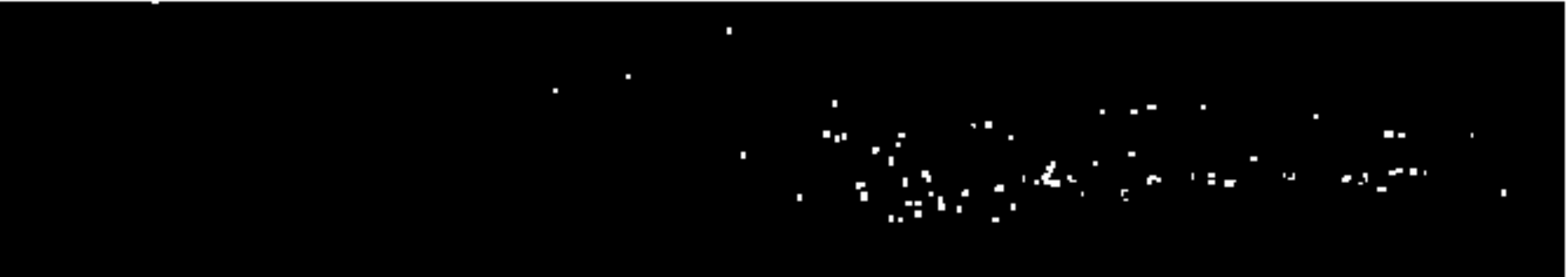}
\caption {Droplets in the \textit{pixel-size class}\ 7-8}
\label{L8}
\end{figure}
\begin{figure} [pt]
\centering
\includegraphics [scale=.75]{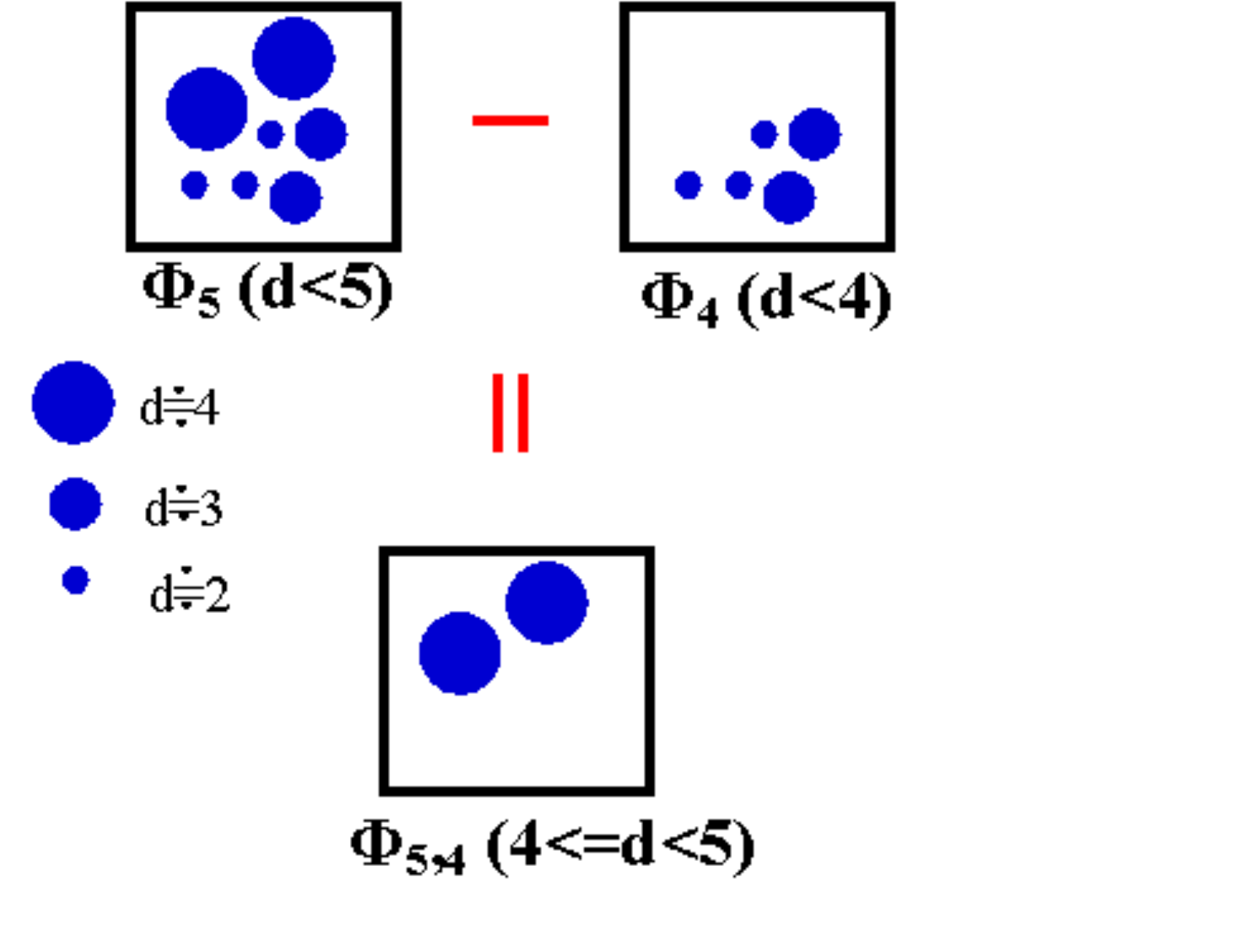}
\caption {How the image $\Psi_{(r,\ r-1)}$ is obtained}
\label{sieve2}
\end{figure}
       After a sequence of images $\Psi_{t}(A)$ , t=1,2,\ldots has
       been
       generated, their complements w.r.t the entire image are
       determined. This sequence is denoted as $\Phi_t(A)$ in
       Figure \ref{sieve2},\ which shows  how the difference between successive
       images of this set can be used to get an image of droplets
       in a particular \textit{pixel-size} class.These images are denoted as $\Phi_{r,r-1}$,
       where r denotes the higher size-class. The \textit{pixel-size classes} are converted to length classes,
        knowing the resolution
       at which the image was captured.

Once this is done, we just have to count the number of droplets in
each such image\ $\Phi_{r,r-1}(A)$ and these values can be
directly used for plotting the droplet size-distribution,\
followed by calculations for the SMD and other diameters.\\
Figure \ref{op4} and Figure \ref{op8} are openings, while Figure
\ref{L8} shows the droplets in the \textit{pixel-size class}\ 7-8
for an image of an actual cryogenic spray.

\subsection{Axiomatic development of granulometric mappings}
Let X be the collection of Euclidean images $\subset \Re^2$.\ Then
a \textit{granulometry} on X is a family of mappings\\
$\Psi_t : X\longmapsto X ,t>0 , $ such that
\begin{enumerate}
\item $\Psi_t(A)\subset A \ \forall \ t>0$ \ldots $\Psi_t$ is
antiextensive.
\item $A\subset B \ \Longrightarrow \Psi_t(A)\subset
\Psi_t(B)$\ldots $\Psi_t$ is increasing.
\item $\Psi_t\circ \Psi_{\acute{t}}=$ $\Psi_{\acute{t}}\circ
\Psi_t=\Psi_{max(t,\acute{t})}, \ \forall \ t,\  \acute{t} >0$.\\
In particular, $\Psi_0(A)=A$.Once we have these axioms, several
results can be easily
derived.One such result is:\\
 For a granulometry ${\Psi_t},$ if $r\geq s, \Psi_r(A) \subset
\Psi_s(A)$\\

If we add the following two  axioms to the existing three, we get
an \textit{Euclidean granulometry} For any $t>0$:
\item  $\Psi_t$ is compatible with translation.
\item For an image A,\ $\Psi_t=t\ \Psi_t(1/t \ A)$

\end{enumerate}
\newpage
\section{Conclusions}
In this report, it is seen how Mathematical Morphology can be used
in  processing the image of a spray to quantify the structures
observed in it.Though this technique was developed, considering
the need for analyzing Cryogenic sprays, it could be applied to
storable propellants, as well.

Morphological opening is however, a shape-distorting operation and
therefore, no attempt must be made to relate the \textit{shapes}
of droplets, in the images comprising the granulometry, with those
in the actual spray.

\end{document}